# A Data Mining view on Class Room Teaching Language

Umesh Kumar Pandey[1], S. Pal[2]

[1] Research Scholar,
Singhania University, Jhunjhunu, Rajasthan, India

[2] Dept. of MCA, VBS Purvanchal University
Jaunpur – 222001, Uttar Pradesh, India

**Abstract**
From ancient period in India, educational institution embarked to use class room teaching. Where a teacher explains the material and students understand and learn the lesson. There is no absolute scale for measuring knowledge but examination score is one scale which shows the performance indicator of students. So it is important that appropriate material is taught but it is vital that while teaching which language is chosen, class notes must be prepared and attendance. This study analyses the impact of language on the presence of students in class room. The main idea is to find out the support, confidence and interestingness level for appropriate language and attendance in the classroom. For this purpose association rule is used.

## 1. Introduction

Higher education has attained a key position in the knowledge society under globalize economy [12].

Students' academic performance hinges on diverse factors like personal, socio-economic, psychological and other environmental variables. The prediction of students with high accuracy is beneficial to identify the students with low academic achievement [14].

Data mining involves many different tasks. So it is a complex process but it has a high degree of accuracy [9].

An association rule data mining model identifies specific types of data association. Association rule mining finds all rules that satisfy some minimum support and confidence constraint that it is the target of on mining is not predetermined [4].

Before the globalization most of the education program run in Hindi or in regional language. But after the globalization of Indian economy several professional courses started in English language. Students are bound to study these courses in English language. The real problem arises in the classroom i.e. which language a teacher must use to communicate knowledge to student because in a day long students converses in Hindi or any other regional language.

In this paper it is tried to find out the association of language and attendance.

## 2. Background and related work

Due to high accuracy and prediction quality data mining technique is widely used in different areas. Education sector is also enriched with the help of this technique. A number of journal and literature are available containing educational data mining. Few of them are listed below for reference.

Merceron A et al. concluded that association technique requires not only that adequate thresholds be chosen for the two standard parameters of support and confidence, but also that appropriate measures of interestingness be considered to retain meaning rules that filter uninteresting ness ones out[10].

Tan P N, kumar V and srivastava J proposed that many measures provide conflicting information provide the interestingness of a pattern and the best metric to use for a given application domain is rarely known. They made a comparative study on all interestingness measures and present a small set of table such that an expert can select a desirable measure [16].

Merceron A et al. investigated the interestingness of the association rules found in the data to look for mistakes often made together while solving an exercise, and found strong rules associating three specific mistakes [11].

Oladipupo O.O. and Oyelade O.J. study has bridge the gap in educational data analysis and shows the potential of the association rule mining algorithm for enhancing the effectiveness of academic planners and level advisers in higher institutions of learning [13].

Hijazi and Naqvi conducted study to analyze "students' attitude towards attendance in class, hours spent in study on daily basis after colleges, family income, students' mother's age and mother's education are significantly related with student performance [7].

Pandey U. K. and Pal S. conducted study on the student performance based by selecting 600 students from different colleges of Dr. R. M. L. Awadh University, Faizabad, India. By means of Bayes Classification on category, language and background qualification, it was found that whether new comer students will performer or not [19].

Bray conducted a comparative study to find out private tutoring percentage in different countries. He revealed that India has highest percentage of student's compare to Malaysia, Singapore, Japan, China and Sri Lanka [3].

Al Raddaideh et al. applied, decision tree model to predict the final grade of students, who studied the C++ course [1].

## 3. Data Mining

The research in database and information technology has given rise to an approach to store and manipulate these data for further decision making [2]

Data mining is often set in the broader context of knowledge discovery in database or KDD. The KDD process involves several stages: selecting the target, processing data, transforming them if necessary, performing data mining to extract pattern and relationship and then interpreting and assessing the discovered structures [5].

Han and kamber describes data mining software that allows the users to analyze data from different dimensions and categorize it [6].

Higher education institutions face lack of deep and adequate knowledge which is root cause for low achievement of defined quality objectives. This gap can be bridged by using data mining technique. It will fill up the gaps found in educational environment.

Dunham [9] categorized various models and tasks of data mining into two groups: predictive and descriptive. Figure 1 shows most commonly used data mining models and tasks:

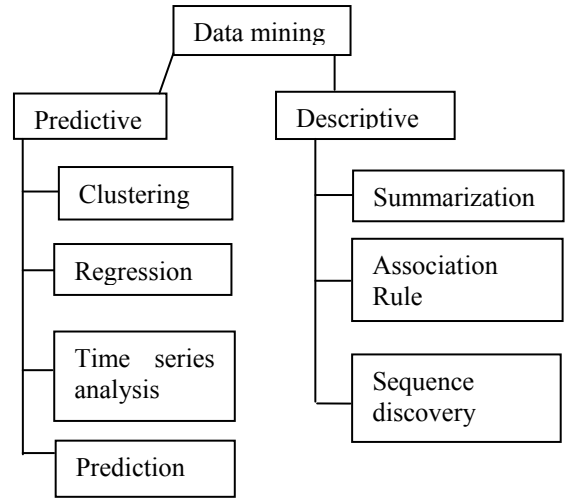

Figure 1: Data Mining Models

## 4. Association rule

One of the popular descriptive data mining is association rule owing to its extensive use in marketing and retail communities in addition to many other diverse fields [18]. Association rule mining is one of the important technique which aims at extracting, interesting correlation, frequent patterns, associations or casual structures among set of items in the transaction databases or other data mining repositories [17].

Association rules are used to show the relationships between item sets. An association rule is accompanied with two measures support and confidence. These terms can be defined as:

**Association rule:** Given a set of items $I=\{I_1, I_2, \ldots I_m\}$ and a database of transaction $D=\{t_1, t_2, \ldots t_n\}$ where $t_i=(I_{i1}, I_{i2}, \ldots \ldots I_{ik}\}$ and $I_{ij} \in I$, an association rule is an implication of the form $X \Rightarrow Y$ where $X, Y \subset I$ are sets of items called item sets and $X \cap Y = \phi$ [9].

It is a fact that strong association rules are not necessarily interesting [16]. Following table has list of interestingness measures for association rule:

| # | Measure | Definition |
|---|---------|-----------|
| 1 | φ-coefficient | $\frac{P(A,B) - P(A)P(B)}{\sqrt{P(A)P(B)(1-P(A))(1-P(B))}}$ |
| 2 | Goodman-Kruskal's (λ) | $\frac{\sum_j \max_k P(A_j, B_k) + \sum_k \max_j P(A_j, B_k) - \max_j P(A_j) - \max_k P(B_k)}{2 - \max_j P(A_j) - \max_k P(B_k)}$ |
| 3 | Odds ratio (α) | $\frac{P(A,B)P(\overline{A},\overline{B})}{P(A,\overline{B})P(\overline{A},B)}$ |
| 4 | Yule's Q | $\frac{P(A,B)P(\overline{A}\overline{B}) - P(A,\overline{B})P(\overline{A},B)}{P(A,B)P(\overline{A}\overline{B}) + P(A,\overline{B})P(\overline{A},B)} = \frac{\alpha-1}{\alpha+1}$ |
| 5 | Yule's Y | $\frac{\sqrt{P(A,B)P(\overline{A}\overline{B})} - \sqrt{P(A,\overline{B})P(\overline{A},B)}}{\sqrt{P(A,B)P(\overline{A}\overline{B})} + \sqrt{P(A,\overline{B})P(\overline{A},B)}} = \frac{\sqrt{\alpha}-1}{\sqrt{\alpha}+1}$ |
| 6 | Kappa (κ) | $\frac{P(A,B) + P(\overline{A},\overline{B}) - P(A)P(B) - P(\overline{A})P(\overline{B})}{1 - P(A)P(B) - P(\overline{A})P(\overline{B})}$ |
| 7 | Mutual Information (M) | $\frac{\sum_i \sum_j P(A_i, B_j) \log \frac{P(A_i, B_j)}{P(A_i)P(B_j)}}{\min(-\sum_i P(A_i) \log P(A_i), -\sum_j P(B_j) \log P(B_j))}$ |
| 8 | J-Measure (J) | $\max\left(P(A,B) \log(\frac{P(B\|A)}{P(B)}) + P(A\overline{B}) \log(\frac{P(\overline{B}\|A)}{P(\overline{B})}),\right.$ $\left. P(A,B) \log(\frac{P(A\|B)}{P(A)}) + P(\overline{A}B) \log(\frac{P(\overline{A}\|B)}{P(\overline{A})})\right)$ |
| 9 | Gini index (G) | $\max\left(P(A)[P(B\|A)^2 + P(\overline{B}\|A)^2] + P(\overline{A})[P(B\|\overline{A})^2 + P(\overline{B}\|\overline{A})^2]\right.$ $- P(B)^2 - P(\overline{B})^2,$ $P(B)[P(A\|B)^2 + P(\overline{A}\|B)^2] + P(\overline{B})[P(A\|\overline{B})^2 + P(\overline{A}\|\overline{B})^2]$ $\left. - P(A)^2 - P(\overline{A})^2\right)$ |
| 10 | Support (s) | $P(A,B)$ |
| 11 | Confidence (c) | $\max(P(B\|A), P(A\|B))$ |
| 12 | Laplace (L) | $\max\left(\frac{NP(A,B)+1}{NP(A)+2}, \frac{NP(A,B)+1}{NP(B)+2}\right)$ |
| 13 | Conviction (V) | $\max\left(\frac{P(A)P(\overline{B})}{P(A\overline{B})}, \frac{P(B)P(\overline{A})}{P(B\overline{A})}\right)$ |
| 14 | Interest (I) | $\frac{P(A,B)}{P(A)P(B)}$ |
| 15 | cosine (IS) | $\frac{P(A,B)}{\sqrt{P(A)P(B)}}$ |
| 16 | Piatetsky-Shapiro's (PS) | $P(A,B) - P(A)P(B)$ |
| 17 | Certainty factor (F) | $\max\left(\frac{P(B\|A) - P(B)}{1 - P(B)}, \frac{P(A\|B) - P(A)}{1 - P(A)}\right)$ |
| 18 | Added Value (AV) | $\max(P(B\|A) - P(B), P(A\|B) - P(A))$ |
| 19 | Collective strength (S) | $\frac{P(A,B) + P(\overline{A}\overline{B})}{P(A)P(B) + P(\overline{A})P(\overline{B})} \times \frac{1 - P(A)P(B) - P(\overline{A})P(\overline{B})}{1 - P(A,B) - P(\overline{A}\overline{B})}$ |
| 20 | Jaccard (ζ) | $\frac{P(A,B)}{P(A) + P(B) - P(A,B)}$ |
| 21 | Klosgen (K) | $\sqrt{P(A,B)} \max(P(B\|A) - P(B), P(A\|B) - P(A))$ |

Figure 2 interestingness measures for association pattern [16]

## 5. Data mining techniques used in this paper

**Support:** The support (s) for an association rule X⇒Y is the percentage of transaction that contain X∪Y [9].

**Confidence:** The confidence or strength (α) for an association rule X⇒Y is the ratio of the number of transaction that contain X [9].

**Cosine:** Consider two vectors X and Y and the angle they form when they are placed so that their tails coincide. When this angle nears 0°, then cosine nears 1, i.e. the two vectors are very similar: all their coordinates are pair wise the same (or proportional). When this angle is 90°, the two vectors are perpendicular, the most dissimilar, and cosine is 0.

$$Co\sin e(X \Rightarrow Y) = \frac{P(X,Y)}{\sqrt{P(X)*P(Y)}}$$

The usual form that is given for cosine of an association rule is X, Y. The closer cosine (X⇒Y) is to 1, the more transactions containing item X also contain item Y, and vice versa. On the contrary, the closer cosine (X⇒Y) is to 0, the more transactions contain item X without containing item Y, and vice versa. This equality shows that transactions not containing neither item X nor item Y have no influence on the result of Cosine (X⇒Y). This is known as the null-invariant property. Note also that cosine is a symmetric measure [10, 11].

**Added value:** The added value of the rule X→Y is denoted by AV (X ⇒Y) and measures whether the proportion of transactions containing Y among the transactions containing X is greater than the proportion of transactions containing Y among all transactions. Then, only if the probability of finding item Y when item X has been found is greater than the probability of finding item Y at all can we say that X and Y are associated and that X implies Y.

$$AV(X \Rightarrow Y) = conf(X \Rightarrow Y) - P(Y)$$

A positive number indicates that X and Y are related, while a negative number means that the occurrence of X prevents Y from occurring. Added Value is closely related to another well-known measure of interest, the lift [10].

**Lift:**

$$Lift(X \Rightarrow Y) = \frac{conf(X \Rightarrow Y)}{P(Y)}$$

An equivalent definition is: $\frac{P(X,Y)}{P(X)*P(Y)}$. Lift is a symmetric measure. A lift well above 1 indicates a strong correlation between X and Y. A lift around 1 says that P(X, Y) = P(X)*P(Y). In terms of probability, this means that the occurrence of X and the occurrence of Y in the same transaction are independent events, hence X and Y not correlated. It is easy to show that the lift is 1 exactly when added value is 0; the lift is greater than 1 exactly when added value is positive and the lift is below 1 exactly when added value is negative [10, 11].

**Correlation:**

$$Correlation = \frac{P(X,Y) - P(X)*P(Y)}{\sqrt{P(X)*P(Y)*(1 - P(X)*(1 - P(Y))}}$$

Correlation is a symmetric measure. A correlation around 0 indicates that X and Y are not correlated, a negative figure indicates that X and Y are negatively correlated and a positive figure that they a positively correlated. Note that the denominator of the division is positive and smaller than 1. Thus the absolute value | cor (X→Y)| is greater than |P(X, Y)-P(X)P(Y)|. In other words, if the lift is

around 1, correlation can still be significantly different from 0 [11].

**Conviction:**

$$Conviction(X \Rightarrow Y) = \frac{(1 - P(Y))}{(1 - conf(X \Rightarrow Y))}$$

Conviction is not a symmetric measure. A conviction around 1 says that X and Y are independent; while conviction is infinite as conf (X→Y) is tending to 1. Note that if P(Y) is high, 1 − P(Y) is small. In that case, even if conf(X, Y) is strong, conviction (X⇒Y) may be small [10].

## 6. Application

In this study data gathered from a degree college named PSRIET, Ranjeetpur Chilbila Pratapgarh, UP, affiliated with Dr. RMLA University, Faizabad, India from the period august 2010 to January 2011. These data are analyzed using Association rule to find the interestingness of student in opting class teaching language. In order to apply this following steps are performed in sequence:

**6.1 Data set:** The data set used in this study was obtained from a degree college of computer science department of course PGDCA of session 2010-11. Initial size of the data is 60.

**6.2 Data selection and transformation:** In this step data selected from various table to extract the information. College organized the classes in three different languages to attract the student that is Hindi, English, and Mix. Hindi medium class contains 80 percent of the lecture in local language and 20 percent English. English medium class contains more than 90 percent in English. Mix medium class contains English and Hindi in almost equal ratio. Class notes provided in each class only English medium. As the course is available in only English medium college tried to find out the student interestingness in opting classroom teaching language. Following Venn diagram (fig. 3) shows the complete attendance picture of student in different medium class.

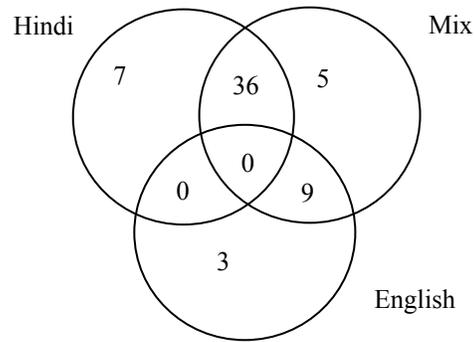

Figure 3: attendance in different medium of classes

So, college tried to find out the student interestingness of class teaching language. In this sequence college focused on the attendance register. It identified that there is some common interest in the student. On the basis of available attendance college prepared the support of each language. Following table 1 shows the support level for each language. It describes that he mix medium class had higher percentage of support. Most of the students participate in this classroom. Lowest percentage of support is for English medium class.

Table 1: Support analysis

| Class medium | Support |
|---|---|
| Hindi | 0.70 |
| English | 0.20 |
| Mix (Hindi & English) | 0.83 |
| {Hindi, Mix} | 0.60 |
| (English, Mix) | 0.15 |

**6.3 Result and analysis:** In this step data selected from various table to extract the information.

Table 2 describes that the most of the students (85.7%) who attend the Hindi medium class also joins the mix medium class because it has highest confidence. 72% of mix medium class students also join the Hindi medium class. 75 % of English medium class students also join the Mix medium class. Only 18 % of mix medium class students joins English medium.

Table 2: Confidence analysis

| Association | Confidence |
|---|---|
| Hindi⇒Mix | 0.857 |
| Mix⇒ Hindi | 0.720 |
| English⇒Mix | 0.750 |
| Mix⇒ English | 0.180 |

Table 3 shows the cosine analysis valve. It is a symmetric analysis. It means two sets give same results in either direction. In this research paper it shows the angular value between two different medium of class. Table shows that Hindi and Mix medium class has lower angle (38.216) in comparison to English and Mix medium. It can be concluded as the Hindi and Mix medium class has more similarity of student than the English and Mix medium class.

Table 3: Cosine analysis

| Association | Cosine | Angle |
|---|---|---|
| Hindi⇒Mix | 0.786 | 38.210 |
| Mix⇒ Hindi | 0.786 | 38.210 |
| English⇒Mix | 0.367 | 68.416 |
| Mix⇒ English | 0.367 | 68.416 |

Table 4 shows the added value analysis. In this table Hindi⇒Mix and Mix⇒ Hindi has positive number which shows that they are related to each other. Occurrence of Hindi does not prevent Mix from occurring similarly occurrence of Mix does not prevent Hindi from occurring. English⇒Mix and Mix⇒ English has negative number which shows that the occurrence of English prevents occurring of Mix similarly occurrence of Mix prevents occurrence of English.

Table 4: Added value analysis

| Association | Added value |
|---|---|
| Hindi⇒Mix | 0.024 |
| Mix⇒ Hindi | 0.020 |
| English⇒Mix | -0.083 |
| Mix⇒ English | -0.020 |

Table 5 contains lift analysis. It is a symmetric analysis. It shows the occurrence of one item to another item. In this table Hindi⇒Mix and Mix⇒Hindi relation has similar positive value (1.029) grater than 1 which shows that occurrence of first is strongly correlated with the other. In the case of English⇒Mix and Mix⇒ English, it has also same positive value but less than 1 which shows that they are negatively correlated.

Table 5: Lift analysis

| Association | Lift |
|---|---|
| Hindi⇒Mix | 1.029 |
| Mix⇒ Hindi | 1.029 |
| English⇒Mix | 0.900 |
| Mix⇒ English | 0.900 |

Table 6 contains correlation value. In this table Hindi⇒Mix and Mix⇒Hindi has similar positive value and English⇒Mix and Mix⇒ English has similar negative value because it is a symmetric measurement. This table shows that is positively correlated whereas English⇒Mix and Mix⇒ English is negatively correlated to each other.

Table 6: Correlation analysis

| Association | Correlation(ϕefficient) |
|---|---|
| Hindi⇒Mix | 0.098 |
| Mix⇒ Hindi | 0.098 |
| English⇒Mix | -0.112 |
| Mix⇒ English | -0.112 |

Table 7 shows conviction analysis. It shows that highest conviction is found in the association of Hindi⇒Mix medium class with value 1.167. Lowest conviction is found in association of English⇒Mix with value 0.667.

Table 7: Conviction analysis

| Association | Conviction |
|---|---|
| Hindi⇒Mix | 1.167 |
| Mix⇒ Hindi | 1.071 |
| English⇒Mix | 0.667 |
| Mix⇒ English | 0.976 |

## 7. Conclusions

Association rules are useful to find the association between two elements and shows interestingness between them. In this paper seven different interestingness parameters are used to find the interestingness between two different medium of class. From the above analysis it can be concluded that the Mix medium class is more preferred over Hindi and English medium class. Another conclusion is extracted from confidence, cosine, AV analysis, lift, correlation and conviction analysis is that most of the Hindi medium class students are showing their interest towards Mix medium class as well as English medium class students also has greater interest in Mix medium class. So college has to organize a Mix medium class to keep attendance at high point.

**Umesh Kumar Pandey** is Assistant Professor in the Department of Computer Applications, PSRIET, Pratapgarh, India. He obtained his M.C.A degree from IGNOU (2004) and M.Phil. in Computer Science from PRIST University, Tamilnadu. He is currently doing research in Data Mining and Knowledge Discovery from Singhaniya University, Rajasthan.

**Saurabh Pal** received his M.Sc. (Computer Science) from Allahabad University, UP, India (1996) and obtained his Ph.D. in Mathematics from the Dr. R. M. L. Awadh University, Faizabad (2002). He then joined the Dept. of Computer Application, VBS Purvanchal University, Jaunpur as Lecturer. At present, he is working as Associate Professor of Computer Applications. S. Pal has authored a commendable number of research papers in international/national Conference/journals and also guides research scholars in Computer Science/Applications.  His research interests include Image Processing, Data Mining and Grid computing.